\documentclass[onecolumn,showpacs,preprintnumbers,amsmath,amssymb,a4]{revtex4}
\usepackage{epsfig}
\usepackage{graphicx}
\usepackage{rotating}
\usepackage{amssymb}
\usepackage{amsmath}
\usepackage{subfigure}

\def\beq{\begin{eqnarray}}
\def\eeq{\end{eqnarray}}

\begin{document}

\preprint{cond-mat/XXXXXXX}

\title{Clustering of solutions in hard satisfiability problems}

\author{John Ardelius}

\affiliation{
SICS Swedish Institute of Computer Science AB\\
P. O. Box 1263\\
SE-16429 Kista, Sweden
}%

\author{Erik Aurell}

\affiliation{
Dept. Computational Biology\\
KTH -- Royal Institute of Technology\\
AlbaNova University Center\\
SE-106 91~Stockholm, Sweden
}%

\author{Supriya Krishnamurthy}

\affiliation{
Dept Information Technology and Communication\\
KTH -- Royal Institute of Technology\\
Forum 105\\
SE-164 40 Kista, Sweden
}%

\date{\today}

\begin{abstract}
We study the structure of the solution space and behavior 
of local search methods on random 3-SAT problems close to the SAT/UNSAT transition. 
Using the overlap measure of similarity between different solutions 
found on the same problem instance we show that the solution space 
is shrinking as a function of $\alpha$. 
We consider chains of satisfiability problems, where clauses
are added sequentially. 
In each such chain, the overlap distribution is first
smooth, and then develops a tiered structure, indicating
that the solutions are found in well separated clusters.
On chains of not too large instances, all solutions
are eventually observed to be in only one small cluster
before vanishing.
This condensation transition point is estimated to be
$\alpha_{c}=4.26$. The transition approximately
obeys finite-size scaling with an apparent critical
exponent of about $1.7$.
We compare the solutions found by a local heuristic, ASAT,
and the Survey Propagation algorithm up to $\alpha_{c}$.
\end{abstract}


\maketitle

\section{Introduction}
\label{sect:intro}
Constraint satisfiability problems (CSPs) are ubiquitous
in application areas such as planning, scheduling, product configuration,
automated electronic design and more~\cite{VanHentenryck89,Mackworth77}.
From the theoretical point of view, a problem's 
computational difficulty is 
determined by embedding it in a family of similar problems of increasing size.
If an algorithm is known and the run-time grows at most polynomially in the size 
of the problem, problems belong to the class P and are considered (relatively)
easy. For CSPs that belong to the hardest class of NP-complete problems, 
supposedly no such algorithm can be found~\cite{Selman-P-NP-ref}.

The subject of this investigation is a benchmark CSP, the
random $k$-satisfiability, or random KSAT, problem.
As described below in section~\ref{s:heuristics}, in KSAT
$M$ propositions in $N$ logical variables are given, each
depending on $k$ variables, and the propositions, or clauses, all
have to be satisfied simultaneously. 
There is a close relation between random KSAT and dilute spin glasses,
where the ``energy'' of a configuration is the number of unsatisfied
clauses. The models are dilute because the number of interactions
each variable has with other variables is Poisson distributed 
with finite mean, \text{i.e.}, generally finite (on a large enough problem),
and ``spin glass-like'' because the clauses are random,
so there is a large amount of frustration.
This has received extensive attention from statistical physicists over
more than ten years~\cite{SAT_UNSAT,MZKST1999,HartmannWeigt,MPZ}. 
The direct analogy of the random KSAT problem is then
whether the spin glass model has a ground state of
energy zero, \textit{i.e.} no violated constraints.

Deterministic algorithms
to solve a CSP will definitely find a solution if there is
one, and will answer in a finite time that no solution exists, 
if that is the case. A prime example is the
DPLL algorithm~\cite{DPLL}. Simpler
procedures, called heuristics or stochastic local search
algorithms, typically find a solution considerably faster, 
but may not always find a solution, although one is there.
Several heuristics can be considered non-equilibrium
relaxation processes in an associated spin glass model
\cite{BARTHEL,SemerjianMonasson03}.
RandomWalksat, the algorithm introduced by Papadimitriou
\cite{RandomWalksat},
walks in a flat landscape, where all non-solutions are treated
equally, while Walksat~\cite{walksat,walksat-WWW}, Focused Metropolis Search (FMS)~\cite{FMS} and 
Average SAT (ASAT)~\cite{ArdeliusAurell06}
are sensitive, in different ways, to the number of violated
constraints.

A main prediction of statistical physics on random KSAT problems
has been the existence of a clustering transition below the
SAT/UNSAT transition, a picture which has passed through several
refinements. In the original version the set of solutions was
predicted to be connected in a single cluster below a threshold
which for 3SAT is at about 3.9~\cite{BiroliMonassonWeigt00}. Above this threshold the
set of solutions was supposed to break up into a large set of smaller clusters.
This scenario has been proven rigorously for $K\ge 8$~\cite{MMZ-05},
with the transition value only however determined approximatively.
In a recent contribution it has been argued that that 
for $K$ equal to four or higher, the previously determined 
clustering threshold is but one in a series more 
complex transformations~\cite{Krzakala-etal06}. 

It has been believed that different clusters
of solutions and different local minima of the
number of unsatisfied clauses are separated by 
extensive barriers. This has not so far been
shown rigorously
or by systematic arguments from spin glass theory. 
If however this rather natural assumption would be true, 
then one would except that local heuristics have difficulties
beyond the clustering transition,
leaving the interval up to the SAT/UNSAT transition to more
sophisticated non-local algorithms such as
Survey Propagation~\cite{MPZ,MZ,CLUSTER}.
Numerical experiments has however showed that some heuristics actually
work linearly on average well beyond all so far predicted clustering transitions
on 3SAT~\cite{AurellGordonKirkpatrick04,FMS,ArdeliusAurell06}.

In this paper we present numerical evidence that a
``cluster condensation'' transition occurs in random 3SAT,
and determine its value to be approximately $4.26$.
We cannot exclude that the location in fact coincides with the
SAT/UNSAT transition at $4.267$. 
The transition approximately obeys finite-size scaling,
with an apparent critical exponent of about $1.7$.

This paper is structured as follows: in section~\ref{s:heuristics}
we present the KSAT problem in a little more detail,
and review some solving techniques for satisfiability
problems. In section~\ref{s:overlap} we investigate the clustering
transition, and compare the solutions found by the ASAT heuristic
with those found by the Survey Propagation algorithm. 
In section~\ref{s:discussion} we discuss the practical and theoretical
significance of our results.

\section{The random $k$-satisfiability problem and heuristics}
\label{s:heuristics}
\subsection{The SAT problem}
The satisfiability problem (SAT) is a central problem in theoretical as well as
practical computer science. It is the problem of assigning values to $N$ binary
variables, $x\in \{0,1\}$, given $M$ constraints. Each constraint specifies preferred 
values for $K$ of the variables and is said to be \textit{satisfied} if at
least one variable equals its specification. The whole problem is said to be satisfied 
if all constraints are.

SAT problems belongs to the class of NP-complete problems which means that no known 
algorithm is able to solve a worst case instance in time polynomial in system size. 
Since worst case analysis is hard, and may not be relevant to the typical
behavior, one is often also interested in how well a certain method performs 
on a ensemble of \textit{random} instances of the problem. That is, given $N$ 
variables construct $M$ constraints each containing $K$ randomly chosen variables 
with random specified values $\{0,1\}$. This problem is known as \textit{random K-SAT},
it is NP-complete when $K$ is three or greater,  
and is the problem studied in this paper.

Random K-SAT has been shown to display a \textit{phase transition} when varying
 the ratio of the number of constraints to the number of variables 
$\alpha = \frac{M}{N}$~\cite{hardness,SAT_UNSAT}. Below the transition a generic instance is with 
high probability (that is with 100\% certainty when $M,N\rightarrow\infty$) satisfiable, 
and above it its not. This fact, which extends also to other random CSPs such
as coloring, has given the problem great attention from the theoretical 
physics community. Methods from physics have also predicted a second transition, 
below the SAT/UNSAT one, where the solutions are beginning to cluster into distinct 
region of the state space. This divides the SAT region into a EASY and HARD region. 
Lately even more transitions which divides these phases even further have been
suggested~\cite{MezardPalassiniRivoire05,Krzakala-etal06}. The exact locations of the these clustering transition points on 3SAT and 
details of their nature are still open questions.

\subsection{SAT solving techniques}
As discussed above, KSAT problems for $K$ greater than two are NP-complete,
and no algorithm with guaranteed preformance that does not grow very quickly
in the size of the problem is likely to exist. Due to their great practical
interest, algorithms on CSPs in NP-complete classes, have nevertheless been studied
extensively, see~\cite{KautzSelman07} for a recent review.

In parallel to deterministic algorithms (with exponential worst-case behaviour),
there are well-performing non-deterministic algorithms,
or \textit{heuristics}. The currently best achieving methods for random K-SAT (and some other
satisfiability problems) are either message-passing based (Belief Propagation (BP),
Survey Propagation (SP)) or
local search methods (walksat, FMS, ASAT). The former use structured
iterations of guesses (beliefs) between variables in order to reduce the
space of possible configurations in the search. The latter use only local information
together with some target function as a guide in configuration space.
Currently, the hardest instances of random 3SAT
that can be solved with Survey Propagation have a clause to variable 
ratio of 4.25~\cite{MPZ,MZ,BraunsteinMezardZecchina05}, while  
the best local search methods solve problems up to at least 4.21~\cite{FMS,ArdeliusAurell06}
in linear time, on the average. 

There are no a priori theoretical reasons known to us why either message-passing 
or local search heuristics (or both) cannot be pushed beyond their current bounds.
Indeed, it is well known that vanilla-flavored SP can be improved by
backtracking in the decimation step~\cite{Parisi03}.

\section{The overlap distribution and the chain method}
\label{s:overlap}
The overlap between two configurations 
$S^{i}=[s^{i}_{1}...s^{i}_{n}]$. 
and
$S^{j}=[s^{j}_{1}...s^{j}_{n}]$
is 
\begin{equation}
d(i,j) \equiv \sum_{k=0}^{N}\delta(s^{i}_{k},s^{j}_{k})
\end{equation}
In this section we will study the overlaps between a finite number 
$L$ of solutions found on one instance of 3SAT problem.
$L$ will most of the time be less or much less than $N$,
the number of variables, so that the mutual overlaps can
be arranged in an
\textit{overlap matrix}:
\begin{equation}
D = 
\left( \begin{array}{ccccc}
N & d(1,2) & d(1,3) & ... & d(1,L) \\
d(2,1) & N & d(2,3) & ... & d(2,L) \\
d(3,1) & d(3,2) & N & ... & d(3,L) \\
: & : & : & :& : \\
d(L,1) & d(L,2) & d(L,3) & ... & N
\end{array} \right)
\end{equation}
The sorted set of all unique distances 
(self overlaps left out) gives a characteristic sample of the distribution of overlaps 
between solutions found by a particular random algorithm on this instance.
The solutions are found using ASAT~\cite{ArdeliusAurell06} with all starting configurations chosen
independently and randomly. 

To examine how the overlap distribution changes with the number of constraints in the problem 
one starts with a problem instance without any constraints ($\alpha=0$). 
The distribution of overlaps in $D_{\alpha=0}$ are then trivially found to be randomly 
distributed around $N/2$. That is, there are no correlations between different solutions
One then adds constraints one at the the time 
until no solutions can be found, and thereby generate a chain of instances:
\begin{equation}
{\cal C} =  [{\cal I}_{\alpha=0},\ldots,{\cal I}_{\alpha=\alpha_{max}}]
\end{equation}
Let us note that the achievable $\alpha_{max}$ depends on the algorithm, the size of the problem
and the chain. We will first study the overlap distribution in one chain, 
and then compare different chains, and how their properties change with $N$.

\begin{figure}[tbp]
\centering 
  \subfigure[Mean value of overlap]{\label{fig:mean}\includegraphics[scale=0.6]{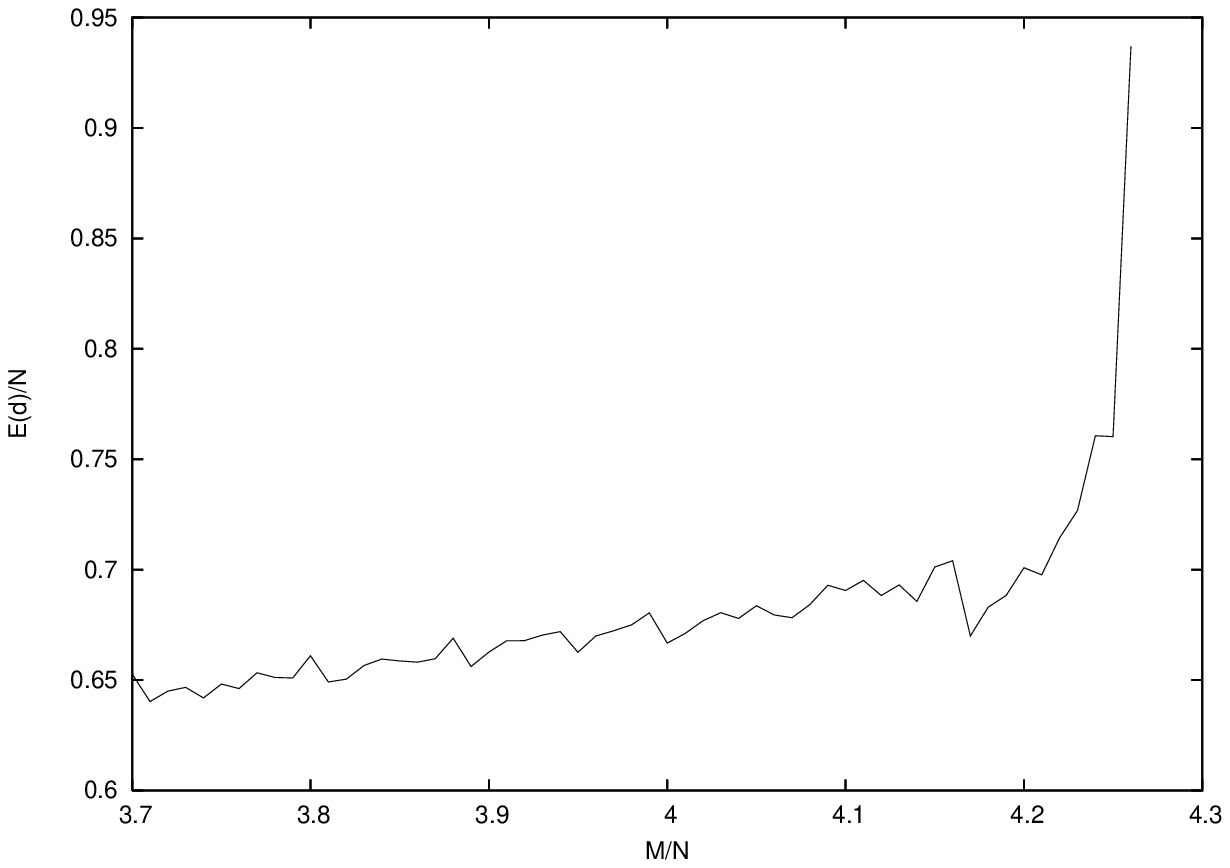}}
  \subfigure[Variance of overlap]{\label{fig:var}\includegraphics[scale=0.6]{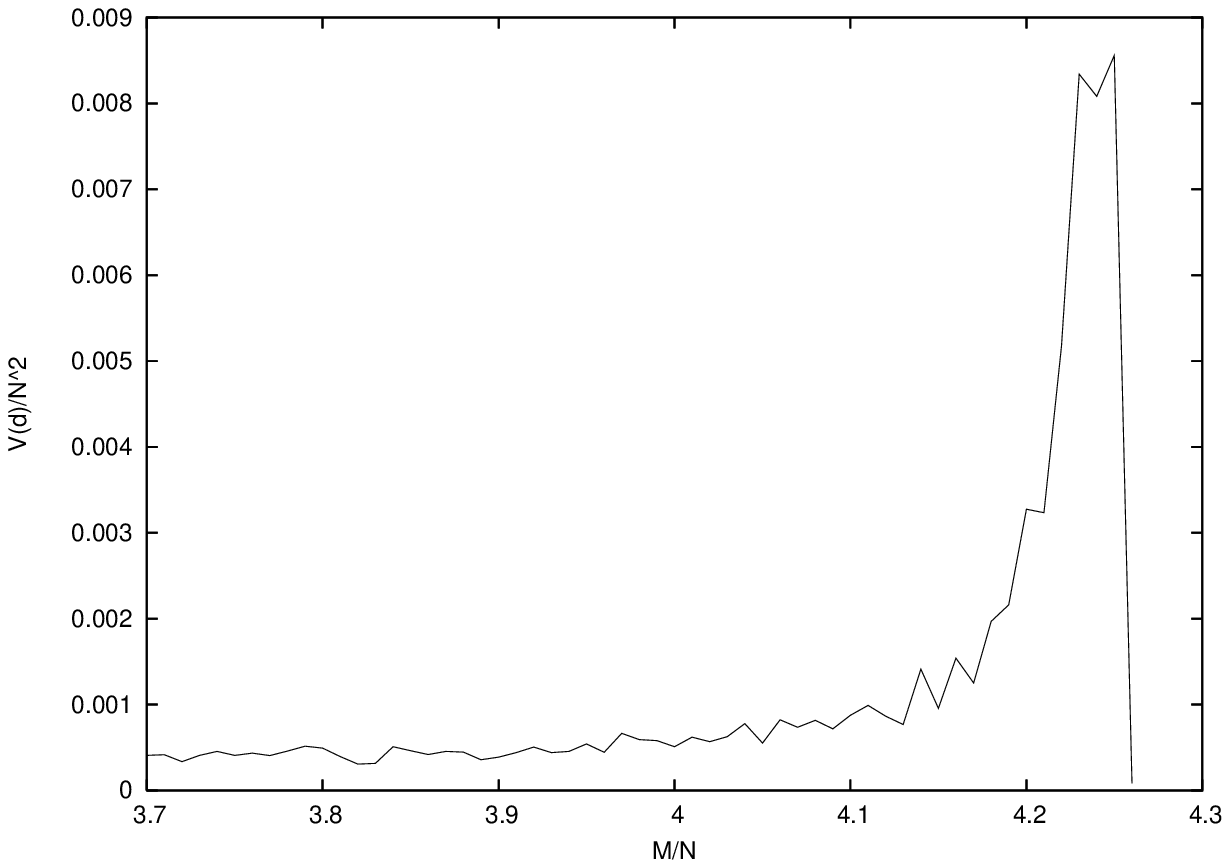}} \\

  \caption{The mean value and the variance of the overlaps for one chain from $\alpha=3.7$ to $\alpha=4.3$. N=2000 variables.}
  \label{fig:stats2k}
\end{figure}

The first property to study is the average of the overlaps between solutions.
Fig.~\ref{fig:stats2k}~a) shows one chain with $N=2000$ variables and $L=40$ solutions
found on each instance. The average clearly increases with
$\alpha$, first smoothly to about $70\%$,
and then sharply to above $90\%$.
Fig.~\ref{fig:stats2k}~b) further shows the variance of the overlaps. This quantity
first also increases, but then eventually decreases to zero.
Taken together, these data imply that the overlap distribution 
starting from $\alpha=3.7$ is first
fairly concentrated at a value of about two thirds, then
develops a fraction of large overlaps, so that both
the average and the variance increases, and finally all
concentrates into the large overlap phase. 

\begin{figure}
    \centering
    \includegraphics{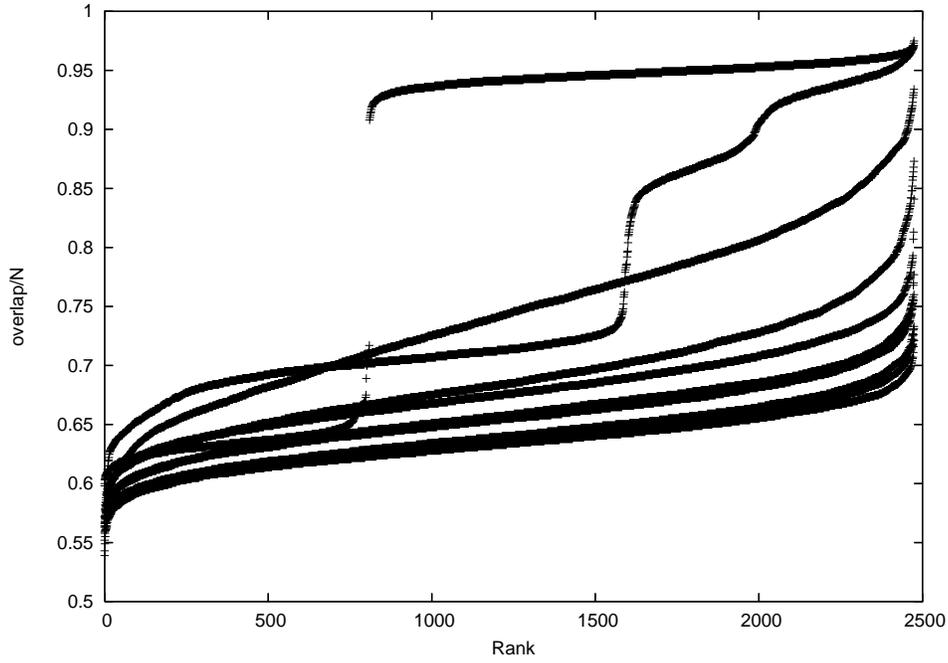}
    \caption{Rank plot of the values of the overlap matrix $D$ for $\alpha=$3.5, 3.6, 3.7, 3.8, 3.9, 4.0, 4.1, 4.2, 4.25 and 4.3. Self-overlaps are left out. For low values of $\alpha$, the curves are continuously distributed around one mean value. When the number of constraints increases the distribution tends to spread out. Increasing $\alpha$ even further results in a break-up of the distribution. The solutions found are then grouped together in clusters. The overlap of solutions in the same cluster is high ($>$0.9N), while the overlap between solutions in different clusters is lower ($\simeq$0.6N)}
    \label{fig:overlap}
\end{figure}

A more detailed picture is obtained by plotting directly the overlap distribution in a chain.
Fig~\ref{fig:overlap} shows that indeed the distribution starts out fairly concentrated, then
grows both higher on average (solutions closer together on average) and also steeper
(some solutions much more similar). At around $4.25$ an interesting transition takes
place, where some solutions remain far apart, some are found much closer. At the last
value investigated, pairs of solutions are apparently either quite close (overlap
above $90\%$) or fairly far apart (overlap around $60\%$). 

The overlap plot of figure \ref{fig:overlap} do not distinguish between
whether there is only one cluster where all high overlap solutions are found 
or a set of clusters. Two possible scenarios that would generate a curve like the one for $\alpha=4.3$ are:
\begin{enumerate}
\item Most of the solutions are found in a small region of the state space. The rest are found randomly distributed.\\
\item All solutions are found in small but separate regions. Few solutions are randomly distributed. \\
\end{enumerate}

In trying to distinguish between the one and many cluster scenarios one can use some kind of clustering detection
 algorithm. We used a simple algorithm described in pseudo code in the following listing.

\begin{verbatim}
1  For a given instance generate n solutions
2  
3  for radius = 0 to N/2
4      rings = 0
5      for i = 0 to n
6          if solution(i) is not inside a ring
7              place a ring with radius centered on solution(i)
8              rings = rings + 1
9          else
10             solution(i) belongs to the covering ring
11         end
12     plot rings vs. radius
13     end
14 end

\end{verbatim}

\begin{figure}
    \centering
    \includegraphics{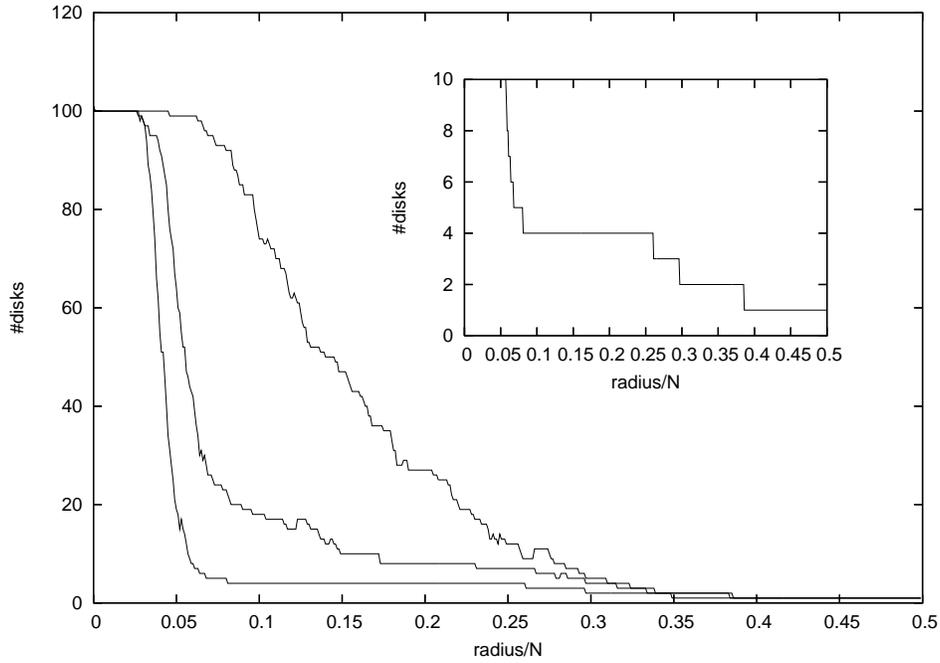}
    \caption{The result of the clustering algorithm shows how many disks with given radius are required to cover the whole set of solutions. The three curves shows this for $\alpha=$4.2, 4.25 and 4.3. The small subplot is an enlargement of the curve for $\alpha=$4.3.}
    \label{fig:clusters}
\end{figure}

This will count the number of rings with radius $r$ that is needed to cover the whole set of solutions. 
The result of the algorithm for values around the clustering transition ($\alpha=$4.3, 4.25 and 4.2) 
for the same instances as in fig \ref{fig:overlap} are depicted in fig~\ref{fig:clusters}.

Comparing the results for the three values of $\alpha$ shows clearly that before the transition when increasing 
the radius of the rings a continuously growing number of solutions are covered. This indicates a (nearly) homogeneous 
density of solutions in the interval from $0.1N$ to $0.3N$. When increasing the number of constraints the 
number of rings needed decreases rapidly around $r=0.05N$ and only a few ($\simeq 10$) rings with radius 
$0.15N$ are required to cover the whole set. In the last instance, 
for $\alpha=4.3$, the number of rings decreases very rapidly from $\sim$ N to only $4$ at radius 
$0.05N$. This indicates that with $4$ rings, each covering about $0.1N$ of the variables, 
the whole set of solutions can be covered, while only one ring is sufficient when the radius is $\sim$ 0.35N. 
The natural interpretation is of course that in this example there are
four different clusters with diameter $0.1N$ which are separated by a distance of $\sim$ 0.3N.
Altogether, the cluster plots favor the second scenario sketched above. 
These facts could not be obtained be the overlap plots alone.\\

\begin{figure}[tbp]
\centering 
  \subfigure[SP/ASAT]{\label{fig:mean}\includegraphics[scale=0.6]{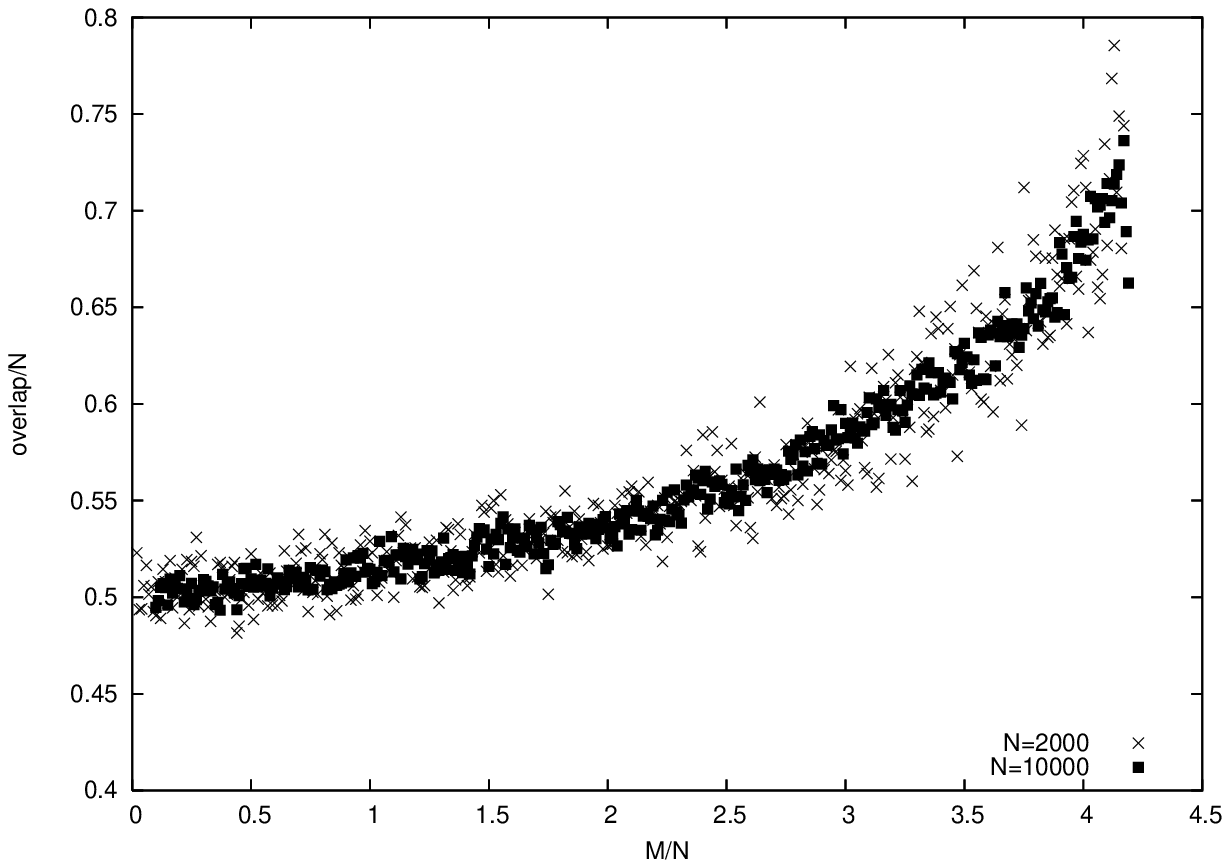}}
  \subfigure[ASAT/ASAT]{\label{fig:var}\includegraphics[scale=0.6]{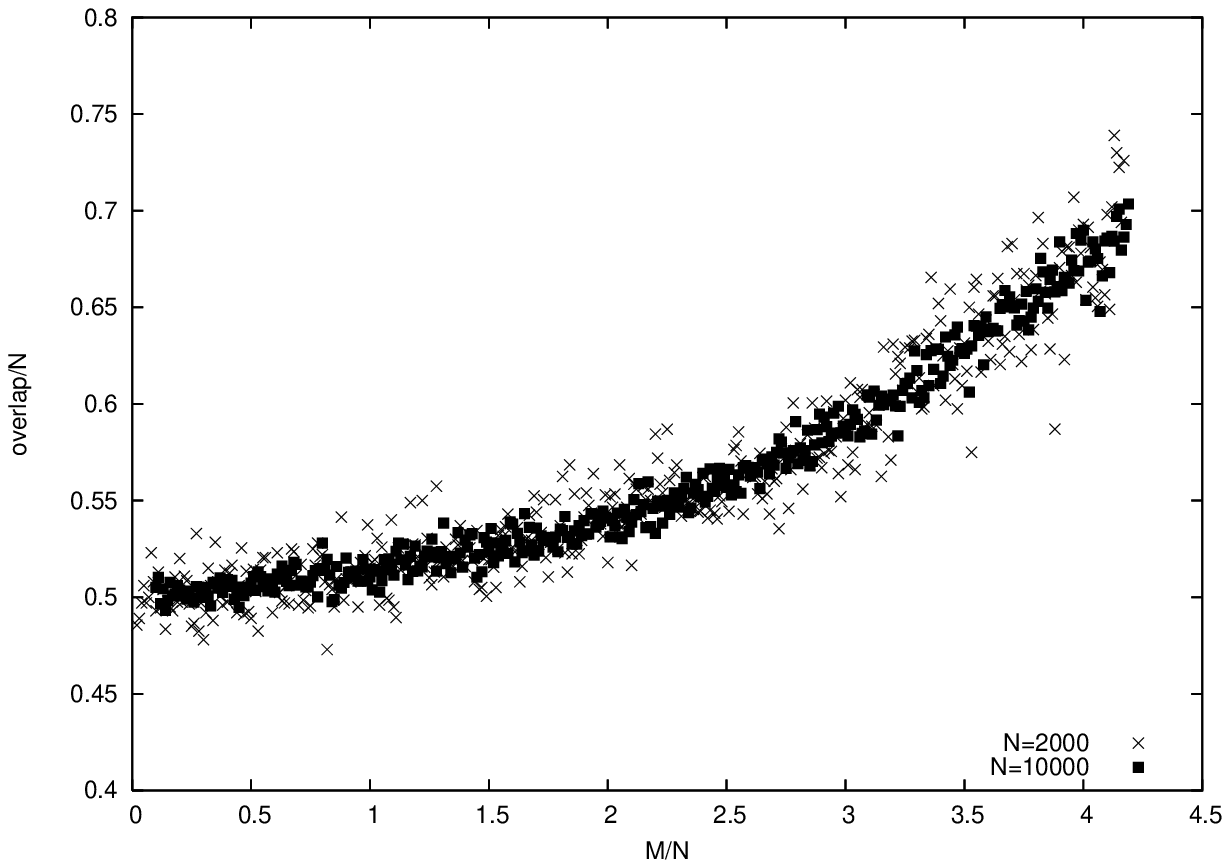}} \\

  \caption{The first plot shows overlap between a solution found by SP and one found by ASAT for $\alpha=$0 to 4.3. The other plot show the overlaps between two different solutions found by ASAT on the same instance. 
The number of variables are 2000 and 10000 in both cases.
}
  \label{fig:spnet}
\end{figure}

An important issue when trying to characterize a set of solutions 
from a finite sample is possible bias in the sampling method.
We cannot answer this question definitely,
but we can compare with other solution methods. To this end 
we have compared ASAT with the Survey Propagation (SP) method~\cite{MPZ,MZ}.
SP, as a satisfiability solver, contains a deterministic step, where
variables are eliminated incrementally based on clues from 
the Survey Propagation message-passing algorithm, which
 for KSAT can be viewed as Belief Propagation in a space of
three values per variable (set 0, set 1, free)~\cite{BraunsteinZecchina04,ManevaMosselWainwright05}.
This deterministic step is followed by a stochastic local
search, by default Selman-Kautz-Cohen walksat~\cite{walksat,walksat-WWW}.
For our purposes, SP can therefore be thought of as practically a deterministic algorithm,
since the solutions found after the stochastic local search typically 
have large overlap.
Fig.~\ref{fig:spnet} 
shows chains of overlaps between solutions found by SP and ASAT
respectively, and two solutions found by ASAT. 
These figures suggest that the solution found by SP is
typical with respect to the solutions found by ASAT, and
hence that the set of solutions sampled by ASAT are relevant
also for very different solution methods.

\begin{figure}[tbp]
\centering 
  \subfigure[]{\label{fig:cpivot}\includegraphics[scale=0.6]{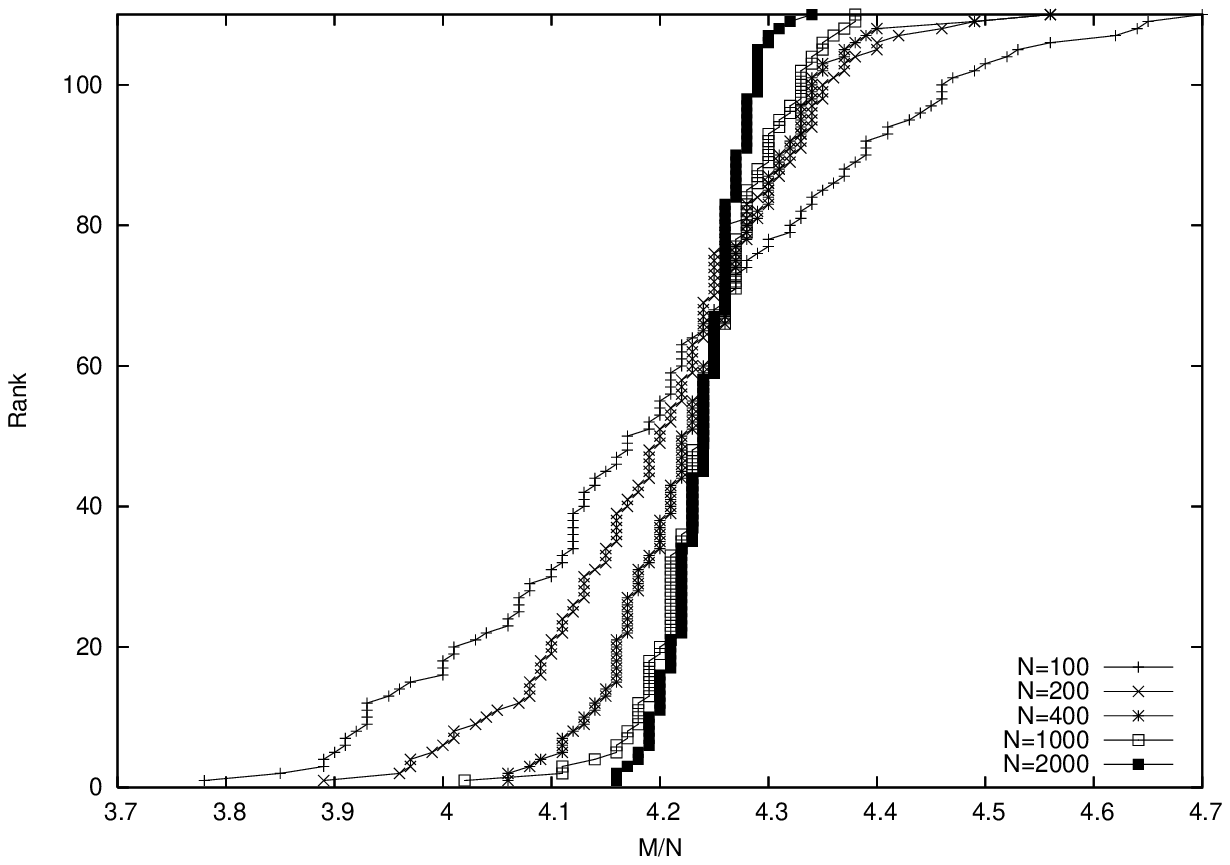}}
  \subfigure[]{\label{fig:scale}\includegraphics[scale=0.6]{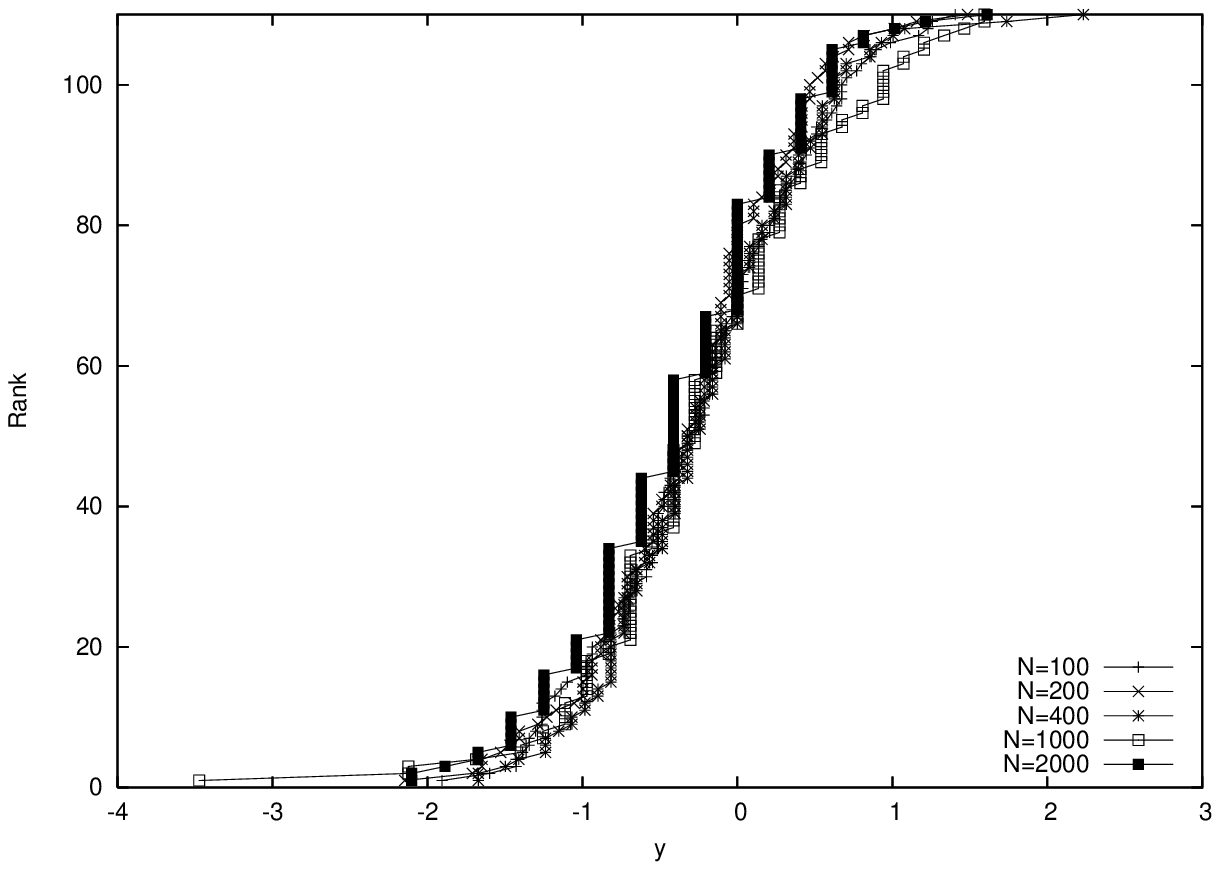}} \\

  \caption{(a) For each chain the point in $\alpha$ where the smallest overlap in the overlap matrix is above $0.8N$ is marked for N=100, 200, 400, 1000 and 2000 variables. 110 chains are used for each N. The point of the jump of each chain marks one point in the rank plot. The mean value of the N=2000 curve is 4.245.(b) Finite size scaling applied to the same data. The best fit is achieved for $\nu=1.7$ and $\alpha_{\infty}=4.26$. The ruggedness of the curves for large N is due to discretization in $\alpha$.
}
  \label{fig:onecluster}
\end{figure}

Finally, we turn to comparing clustering in different chains.
The transition from a finite number of clusters to only one is 
relatively easy to characterize since the both the variance and the minimal value of the overlap matrix $D$ 
change abruptly. Figure \ref{fig:cpivot} shows a rank plot of the point where all solutions found 
have overlap more then $0.8N$. This value gets sharper with larger N which suggests that when 
$N\rightarrow\infty$ a phase transition from several to one cluster takes place. 
Straight-forward finite-size scaling of the data gives a transition at $\alpha=4.26$
with a critical exponent of $1.7$.

\begin{figure}
    \centering
    \includegraphics{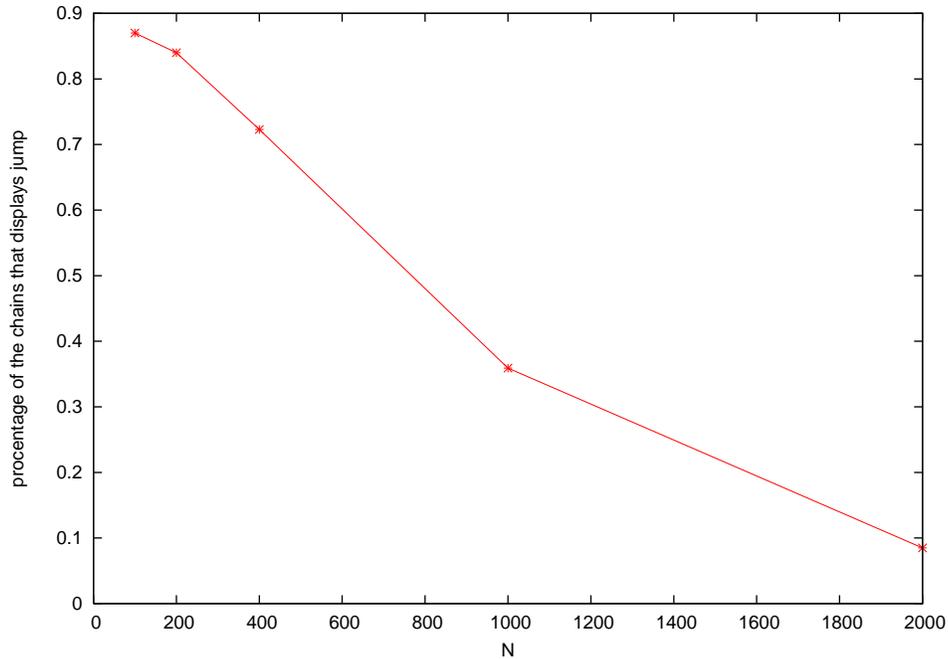}
    \caption{The fraction of instances that display a condensation transition within the time cutoff.}
    \label{fig:success}
\end{figure}

We note that these values are given for the chains that could be prolonged until the clustering transition.
For some instances the average value increases to but does not reach $0.7N$ until no more solutions are found 
by the algorithm, within the time cut-off used here. 
The number of instances that displays a transition before the time cutoff is shown in figure \ref{fig:success}.

\section{Conclusion and discussion}
\label{s:discussion}
The claim that the geometry of the
set of solutions to large random Constraint Satisfaction Problems
is systematically different depending on the structure of the problem has
received wide attention in both statistical physics and computer science
\cite{HartmannWeigt,TCS265,GomesSelman05,AchlioptasNaorPeres05}.
For random $k$-satisfiability problem, rigorous results
exist for $k\ge 8$, new predictions have been recently made
for  $k\ge 4$, while the benchmark random 3SAT problem is now
considered open from the theoretical point of view.

We have determined the clustering structure in numerical experiments, by
solving many 3SAT instances of the same type many times with a simple
stochastic local search procedure ASAT. This heuristic solves 
random 3SAT in linear time, on average, up to at least $4.21$
clauses per variables, as we have shown previously~\cite{ArdeliusAurell06}.
For problem sizes of thousands and tens of thousands of variables it
can also be readily used beyond the linear regime, up to around
the satisfiability/unsatisfiability threshold at about $\alpha_{cr}=4.267$ clauses
per variables.

We find that there is indeed a cluster condensation transition in the solutions
found by ASAT at around $4.26$, very close to $\alpha_{cr}$. The
transition approximately obeys finite-size scaling, with an apparent
critical exponent of about $1.7$. We find indications of the fact that the space of solutions divides in several distinct clusters in some region below the transiontion, approximately between $4.21 < \alpha < 4.26$. We have also
shown that the solutions found by ASAT are compatible with those found
on the same problems by the Survey Propagation (SP) algorithm, in the sense
that the solutions found by SP are not atypical with respect to the
sets of solutions found by ASAT.

We conjecture that the absence of any (observable) clustering 
transition except very close to the satisfiability/unsatisfiability threshold is related
to the surprising effectiveness of simple heuristics in solving large
and hard random 3SAT problems, and perchance other random CSPs as well.

\section*{Acknowledgment}
This work was supported by the Swedish Research Council
(E.A: and S.K) and through IST/FET project EVERGROW 
by the European Union. We thank
Mikko Alava, Petteri Kaski, Pekka Orponen and Sakari Seitz for many
discussions, and Marc M\'ezard and Lenka Zdeborov\'a for useful remarks.

\bibliographystyle{plain} 
\bibliography{sat.bib}

\end{document}